\documentclass{raa}            

\usepackage{graphicx,times}             

\def\la{\mathrel{\mathchoice {\vcenter{\offinterlineskip\halign{\hfil
$\displaystyle##$\hfil\cr<\cr\sim\cr}}}
{\vcenter{\offinterlineskip\halign{\hfil$\textstyle##$\hfil\cr
<\cr\sim\cr}}}
{\vcenter{\offinterlineskip\halign{\hfil$\scriptstyle##$\hfil\cr
<\cr\sim\cr}}}
{\vcenter{\offinterlineskip\halign{\hfil$\scriptscriptstyle##$\hfil\cr
<\cr\sim\cr}}}}}
\def\ga{\mathrel{\mathchoice {\vcenter{\offinterlineskip\halign{\hfil
$\displaystyle##$\hfil\cr>\cr\sim\cr}}}
{\vcenter{\offinterlineskip\halign{\hfil$\textstyle##$\hfil\cr
>\cr\sim\cr}}}
{\vcenter{\offinterlineskip\halign{\hfil$\scriptstyle##$\hfil\cr
>\cr\sim\cr}}}
{\vcenter{\offinterlineskip\halign{\hfil$\scriptscriptstyle##$\hfil\cr
>\cr\sim\cr}}}}}

\def\utw{\smash{\rlap{\lower5pt\hbox{$\sim$}}}}
\def\udtw{\smash{\rlap{\lower6pt\hbox{$\approx$}}}}

\def\diameter{{\ifmmode\mathchoice
{\ooalign{\hfil\hbox{$\displaystyle/$}\hfil\crcr
{\hbox{$\displaystyle\mathchar"20D$}}}}
{\ooalign{\hfil\hbox{$\textstyle/$}\hfil\crcr
{\hbox{$\textstyle\mathchar"20D$}}}}
{\ooalign{\hfil\hbox{$\scriptstyle/$}\hfil\crcr
{\hbox{$\scriptstyle\mathchar"20D$}}}}
{\ooalign{\hfil\hbox{$\scriptscriptstyle/$}\hfil\crcr
{\hbox{$\scriptscriptstyle\mathchar"20D$}}}}
\else{\ooalign{\hfil/\hfil\crcr\mathhexbox20D}}%
\fi}}

\begin{document}  
 \title{ A Search for Evidences of Small-Scale Imhomogeneities in Dense Cores from Line Profile Analysis  
}  

 \volnopage{Vol.0 (200x) No.0, 000--000}      
 \setcounter{page}{1}          

 \author{Lev Pirogov  
    \inst{}  
 }  

 \institute{Institute of Applied Physics, Russian Academy of Sciences,  
                Ulyanova 46, Nizhny Novgorod, Russia;  
            {\it pirogov@appl.sci-nnov.ru}\\  
 }  

 \date{Received~~2017 month day; accepted~~2017~~month day}  

\abstract{  
In order to search for intensity fluctuations on the HCN(1--0) and HCO$^+$(1--0) line profiles which could arise due to possible small-scale inhomogeneous structure long-time observations of the S140 and S199 high-mass star-forming cores were carried out.  
The data were processed by the Fourier filtering method.  
Line temperature fluctuations that exceed noise level were detected.  
Assuming the cores consist of a large number of randomly moving small thermal fragments a total number of fragments is $\sim 4\cdot 10^6$ for the region with linear size $\sim 0.1$~pc in S140 and $\sim 10^6$ for the region with linear size $\sim 0.3$~pc in S199.  
Physical parameters of fragments in S140 were obtained from detailed modeling of the HCN emission in a framework of the clumpy cloud model.  
}  

\authorrunning{L. Pirogov }            
 \titlerunning{A Search for Evidences of Small-Scale Imhomogeneities in Dense Cores }  

 \maketitle  

%
%

\section{Introduction}           
\label{sect:intro}  

The regions where high-mass stars and stellar clusters are born are highly turbulent and inhomogeneous (e.g., Tan et al. \cite{tan14}).  
An extent of turbulence is enhanced in the vicinities of massive stars  
where gas due to various kinds of instabilities 
could fragment into small-scale structures down to the scales unresolved by modern instruments. 
There are many indirect evidences for existence of small-scale unresolved inhomogeneities  
(fragments, clumps) in regions of high-mass star formation.  
This follows from the fact that the observed molecular line profiles
of different species are close to Gaussian profiles without signs of saturation and their widths are much higher than thermal ones (e.g. Kwan \& Sanders \cite{ks86}).
Nearly constant volume densities in clouds  
with strong column density variations (e.g. Bergin et al. \cite{bergin96})  
and detection of C~I emission over large areas correlated with  
molecular maps (e.g. White \& Padman \cite{wp91}, Kamegai et al. \cite{kamegai03}) also imply a small-scale clumpy structure. 

An important evidence for existence of small thermal fragments in high-mass star-forming regions is provided by anomalies of relative intensities of the HCN(1--0) hyperfine components.  
This effect is connected with an overlap of thermally broadened profiles of closely located hyperfine components in the higher HCN rotational transitions  
(mainly, $J$=2--1) and is efficient at kinetic temperatures $\ga 20$~K  
(Guilloteau \& Baudry \cite{gb81}).
Yet, if the local profiles are broadened by microturbulence and are suprathermal as in high-mass star-forming cores ($\ga 2$~km~s$^{-1}$)  
it becomes practically impossible to reproduce the observed HCN(1--0) anomalies  
in a framework of the microturbulent model (Pirogov \cite{pir99}).
In opposite, if the cores consist of small thermal fragments  
with low volume filling factor moving randomly with respect to each other  
the observed HCN(1--0) profiles with intensity anomalies and high linewidths can be easily reproduced (Pirogov \cite{pir99}).

If the observed line profile is a sum of profiles of randomly moving fragments one should expect an existence of intensity fluctuations due to fluctuations of a number of fragments on the line of sight at distinct velocities.  
Martin et al. (\cite{martin}) derived analytical expression for molecular line emission of the cloud consisted of a large number of small identical fragments and Tauber (\cite{tauber96}) obtained an expression for standard deviation of line intensity fluctuations due to such a structure.
Using their approach it is possible to derive from observations parameters of small-scale structure, mainly, a total number of fragments in a telescope beam.  

Previously, we performed long-time observations in various molecular lines (HCN(1--0), CS(2--1), $^{13}$CO(1--0), HCO$^+$(1--0) and some others) of the high-mass star-forming cores which show the HCN(1--0) hyperfine anomalies (S140, S199, S235) and PDR regions (Orion, W3) (Pirogov \& Zinchenko \cite{pz08}, Pirogov et al. \cite{pir12}).
We detected residual fluctuations on line profiles and estimated total number of thermal fragments in a beam using analytical approach. 
By comparing the results of detailed calculations of line emission in a framework of the model consisted of identical thermal fragments (clumpy model) with the observed nearly Gaussian HCN(1--0) profiles, estimates of sizes and densities of fragments were obtained for S140 and S199.
Yet, these results suffered from the drawback connected with arbitrary chosen parameters of the method of extraction residual intensity fluctuations from line profiles.  
In this paper new observational results of higher quality for the S140 and S199 cores in the HCN(1--0) and HCO$^+$(1--0) lines are presented.  
The lines in these objects have nearly Gaussian profiles which is important for comparison with the results of the clumpy model calculations.  
To estimate standard deviations of residual line intensity fluctuations that could be due to small-scale clumpy structure a new Fourier filtering method is used.  
This helped to recalculate parameters of small-scale structure including total number of fragments in the beam for S140 and S199 and physical parameters of fragments for S140.

\section{Analytical Model}  

Considering model cloud consisted of identical randomly moving fragments  
with low volume filling factor and assuming that velocity dispersion  
of fragment motions ($\sigma$) is much higher than inner velocity dispersion  
($v_0$) Martin et al. (\cite{martin}) obtained an expression for cloud's optical depth  
($\tau$) which is proportional to $N_c$,  
the number of fragments in a column with cross-sectional area of single fragment.  
Using this approach for $N_c\la 10$ Tauber (\cite{tauber96}) derived an expression
which can be written as follows:  

\begin{equation}  
\frac{\Delta T_{\rm R}}{T_{\rm R}}=  
\frac{\tau}  
{(e^{\tau}-1)\sqrt{K\,N_{\rm tot} \frac{v_0}{\sigma}}} \hspace{2mm},  
\label{eq:Ntotal}  
\end{equation}  

\noindent{where $\Delta T_{\rm R}$ is a standard deviation of fluctuations of line radiation temperature in some range near line center, $T_{\rm R}$ is a peak line radiation temperature, $K$ is a factor depending on the optical depth distribution within a fragment.}  
For the Gaussian distribution $K$ is equal to 1, for the case of opaque discs it is equal to $\pi$.
Since contribution of emission of an ensemble of small fragments  
is statistically independent from atmospheric and instrumental noise,  
a standard deviation of temperature fluctuations due to small fragments  
can be calculated as:  
$\Delta T_{\rm R}=\sqrt{\Delta T_{\rm L}^2-\Delta T_{\rm N}^2}$,
where $\Delta T_{\rm L}$ and $\Delta T_{\rm N}$ are the observed standard deviations  
of temperature fluctuations within and outside line profile range, respectively.  
Thus, knowing $\Delta T_{\rm R}$, $T_{\rm R}$, kinetic temperature and line optical depth ($\tau$)  
it is possible to estimate a number of thermal fragments in the beam ($N_{\rm tot}$).
Yet, in order to detect radiation temperature fluctuations due to such a structure,  
observations with high signal-to-noise ratio and with high spectral resolution are needed.  
Another problem of this approach is connected with correct measurement of $\Delta T_{\rm R}$.

\section{The Results of Observations}  
\label{sec:observations}  

We carried out observations of two high-mass star-forming cores, S140 and S199, in the HCN(1--0) line at 88.6~GHz with the IRAM-30m telescope in 2010 and in the HCN(1--0) and HCO$^+$(1--0) lines (at 88.6~GHz and 89.2~GHz, respectively) with the OSO-20m telescope in 2017.  
In addition, we observed these sources in the H$^{13}$CN(1--0) and H$^{13}$CO$^+$(1--0) lines with the OSO-20m telescope in 2017.
The IRAM-30m beam at these frequencies is $\sim 29''$, the OSO-20m beam is $\sim 41''$.
System noise temperatures were $\sim 130-180$~K and $\sim 170-240$~K, frequency resolutions were 39~kHz and 19~kHz in the IRAM-30m and the OSO-20m observations, respectively.  
After several hours of integration in the frequency switching mode  
the noise r.m.s. were $\sim 0.01$~K and $\sim 0.02$~K, for the IRAM-30m
and the OSO-20m observations, respectively.  
The observed profiles towards S140 and S199 contain ``quiet'' nearly Gaussian component  
(line widths $\sim 2.5$~km~s$^{-1}$) and high-velocity wing emission of lower amplitude.  
For the purpose of our analysis high-velocity components were subtracted.  
The source coordinates, distances and linear resolutions at $\sim 89$~GHz are given in Table~\ref{sources}.

\begin{table}[hbtp]  
\caption{Source list}  
\smallskip  
\begin{tabular}{lrrcc}\hline\noalign{\smallskip}  
Source          & $\alpha$(2000) & $\delta$(2000) & $D$      & Linear Resolution\\  
                &${\rm (^h)\  (^m)\  (^s)\ }$  
                                 &$(^o)$  $(^{\prime})$  $(^{\prime\prime}$)  
                                                  & (pc)  & (pc)\\  
\noalign{\smallskip}\hline\noalign{\smallskip}  
S140 (L1204)  &  22 19 18.4 & 63 18 45   & 764(27) (Hirota et al. \cite{hirota08}) & $\sim 0.11$ (IRAM-30m) \\  
            &             &            &                              & $\sim 0.15$ (OSO-20m) \\  
S199 (IC1848) &  03 01 32.3 & 60 29 12   & 2200(200) (Lim et al. \cite{lim14}) & $\sim 0.3$ (IRAM-30m)  \\  
\noalign{\smallskip}\hline\noalign{\smallskip}  
\end{tabular}  

\label{sources}  
\end{table}  

\begin{figure}[h]  
\begin{minipage}[t]{0.9\textwidth}  
\centering  
 \includegraphics[width=3cm,angle=-90]{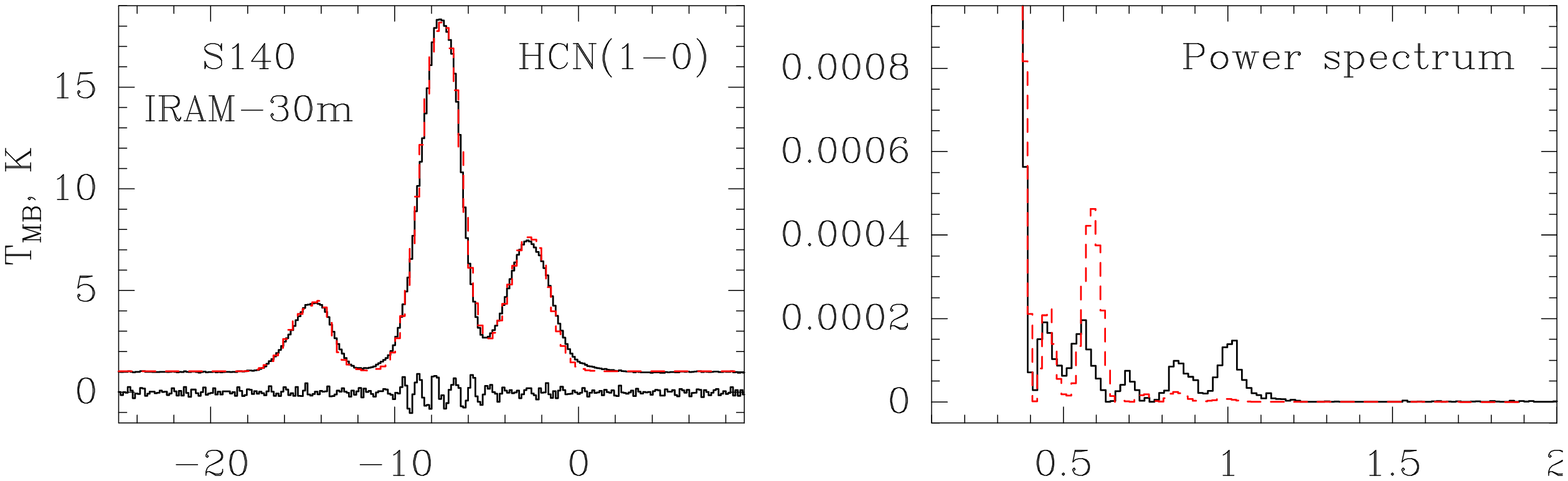}  
\end{minipage}%
\vskip 1mm  
\begin{minipage}[t]{0.9\textwidth}  
\centering  
 \includegraphics[width=3cm,angle=-90]{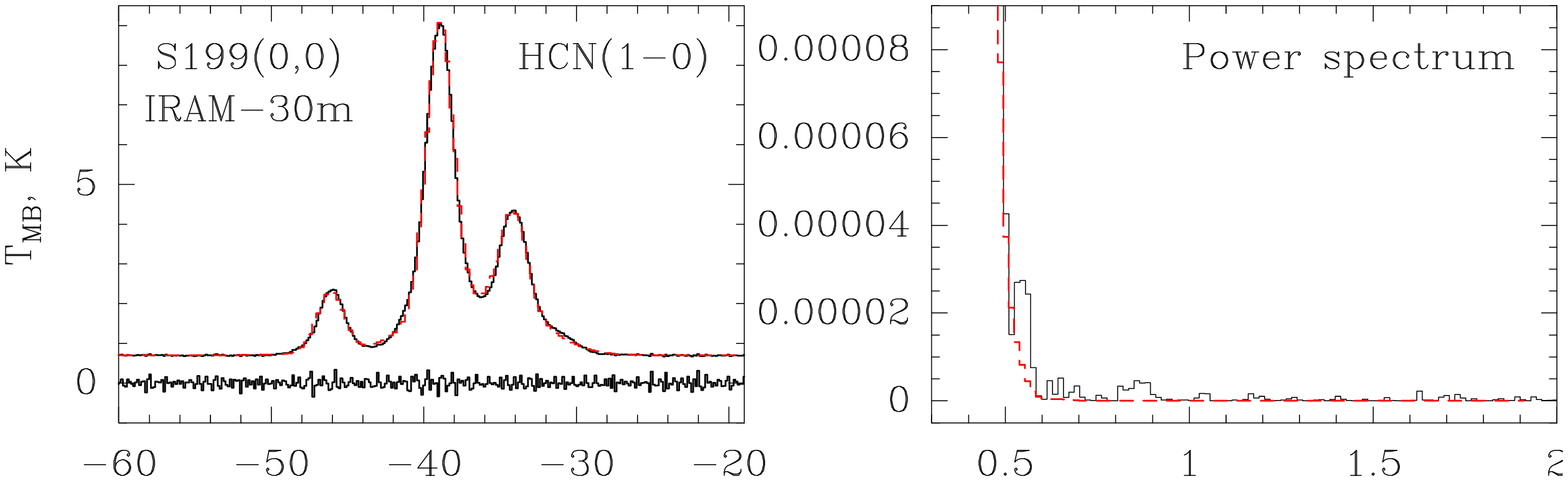}  
\end{minipage}%
\vskip 1mm  
\begin{minipage}[t]{0.9\textwidth}  
\centering  
 \includegraphics[width=3cm,angle=-90]{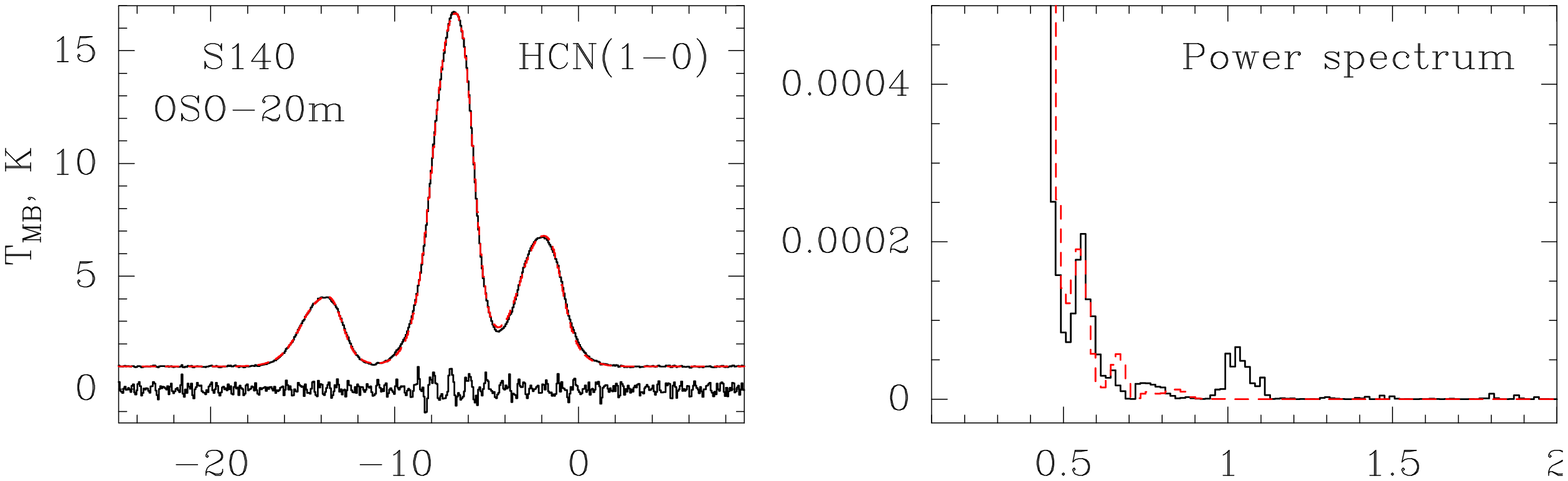}  
\end{minipage}%
\vskip 1mm  
\begin{minipage}[t]{0.9\textwidth}  
\centering  
 \includegraphics[width=3.3cm,angle=-90]{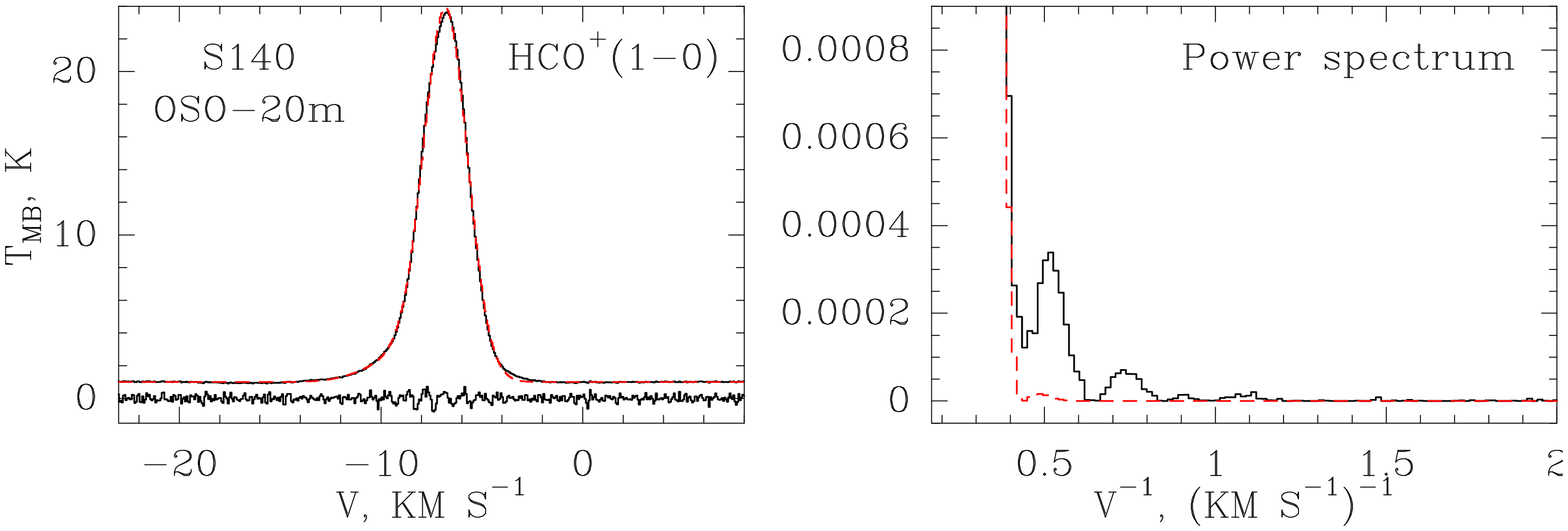}  
\end{minipage}%
\caption{{\small The observed HCN(1--0) and HCO$^+$(1--0) profiles  
in S140 and in S199 
(left panels) and the corresponding power spectra (Fourier transform)  
for low amplitudes (right panels).
Residual noise obtained after filtering power spectra with arbitrary value $F_{\rm eff}$=1~(km~s$^{-1})^{-1}$  
and multiplied by a factor of 10 are shown under the observed profiles.  
Red dashed curves correspond to the fits by overlapping Gaussian  
functions (triplets in the case of HCN(1--0)) and their power spectra.}}  
\label{spectra}  
\end{figure}

In order to estimate standard deviations of radiation temperature fluctuations  
on line profiles that could be due to small-scale structure  
($\Delta T_{\rm R}$) it is necessary to remove correctly the main component from line profiles.  
As was pointed out by Tauber (\cite{tauber96}) one of the possible methods  
is to do Fourier high-pass filtering.  
This method is based on the idea that the small-scale structure should produce  
much broader Fourier (power) spectrum than the main line profile. 
After filtering, 
a noise-like residual spectrum, probably, with different standard deviations  
within and outside line range is obtained.  
Previously (Pirogov \& Zinchenko \cite{pz08}; Pirogov et al. \cite{pir12}) we used this method taking arbitrary filter boundaries to reject harmonics of power spectra corresponding to the main line profile.  

In Fig.~\ref{spectra} the observed profiles and the corresponding power spectra  
are shown on the left and the right panels, respectively.  
The power spectra contain features with low amplitudes at inverse velocities higher than the main Gaussian profile ranges
($> 0.4-0.5$~(km~s$^{-1})^{-1}$) implying small deviations from the Gaussians.  
Fitting of the observed profiles by 2-3 overlapping Gaussians  
(or triplets in the case of HCN) with suprathermal widths it is possible to reproduce some of the low-amplitude features for inverse velocities up to $\sim 0.7$~(km~s$^{-1})^{-1}$ (Fig.~\ref{spectra}, right panels).
The spectral features at higher inverse velocities
could be attributed to the small-scale clumpy structure 
as well as to atmospheric and instrumental noise.

\section{Fourier Filtering and the $\Delta T_{\rm R}(F_{\rm eff})$ Dependencies}  
\label{sec:model}  

\begin{figure}[hbtp]  
\centering  
 \includegraphics[width=55mm,angle=-90]{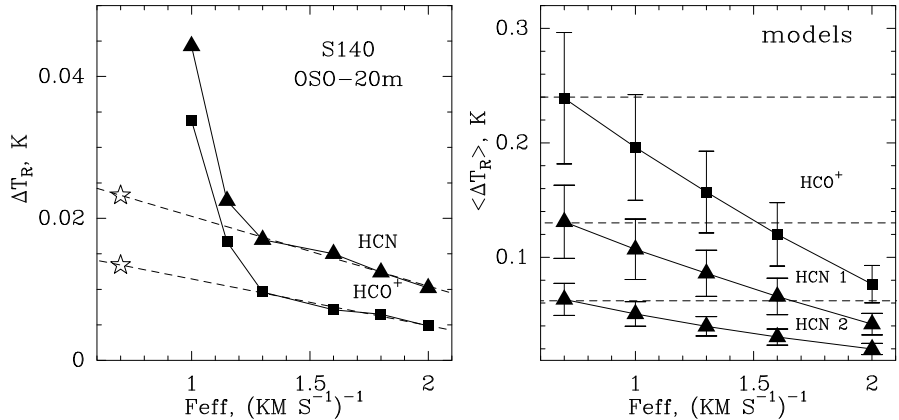}  
\caption{{\small The standard deviations $\Delta T_{\rm R}$ calculated for the  
observed (left) and model (right) HCN(1--0) and HCO$^+$(1--0) profiles  
for different $F_{\rm eff}$ values.  
The model $\Delta T_{\rm R}$ are the mean values of hundred model runs and error bars denote their standard deviations.  
The dashed lines on the right panel correspond to the analytical estimates  
of $\Delta T_{\rm R}$ derived from the equation (\ref{eq:Ntotal}).
Stars on the left panel denote the $\Delta T_{\rm R}$ values taken for calculations of total number of fragments in the beam.}}  
\label{dtr}  
\end{figure}

In order to select an optimal boundary of the high-pass Fourier filter  
($F_{\rm eff}$) the filtering has been done for different values  
of $F_{\rm eff}$ from 0.7 to 2~(km~s$^{-1})^{-1}$  
and the $\Delta T_{\rm R}$ values have been calculated for the 3 km~s$^{-1}$ line range of the observed HCN(1--0) and HCO$^+$(1--0) profiles (23 and 46 velocity channels for the IRAM-30m and OSO-20m data, respectively). 
The results for S140 (OSO-20m) are shown in Fig.~\ref{dtr} (left).
There is a sharp decrease of $\Delta T_{\rm R}$ with increasing $F_{\rm eff}$. 
For $F_{\rm eff}\ga 1.3$~(km~s$^{-1})^{-1}$ the dependencies become nearly linear. 
Similar behavior is found for the IRAM-30m data.  

For comparison we performed test calculations of the HCN and HCO$^+$ excitation  
in the framework of a model cloud consisted of identical thermal fragments  
with small volume filling factors moving randomly with respect to each other with random velocities having the Gaussian distribution.  
The line profile from each fragment is a Gaussian with thermal width. 
A simplified version of the 1D clumpy model described previously  
(Pirogov \cite{pir99}, Appendix; Pirogov \& Zinchenko \cite{pz08}) is used.  
The model matches the conditions of the analytical approach and the model line intensities and widths are close to the observed ones for S140. 
In order to speed up test calculations a number of fragments in the models was reduced. This led to higher values of model $\Delta T_{\rm R}$ values compared with the observed ones. 
 
Varying initial values of random generator one could change spatial distribution and velocities of fragments.  
We performed hundred runs with different initial values of random number generator for two HCN and one HCO$^+$ test models and processed the results in the same way  
as the data of observations, namely, by filtering corresponding power spectra  
for different $F_{\rm eff}$ values and calculating  
$\Delta T_{\rm R}$ for the  3~km~s$^{-1}$ line range.

There is a scatter in the $\Delta T_{\rm R}$ model values from one model run to another.  
For each $F_{\rm eff}$ the mean $\Delta T_{\rm R}$ value and dispersion  
have been calculated and the resulted $\langle\Delta T_{\rm R}\rangle(F_{\rm eff})$ dependencies are plotted in Fig.~\ref{dtr} (right).  
They are close to linear ones and the  
$\langle\Delta T_{\rm R}\rangle$ value at $F_{\rm eff}=0.7$~(km~s$^{-1})^{-1}$ is nearly equal  
to the analytical estimate calculated from the equation (\ref{eq:Ntotal}) for the case of opaque discs.  
The $\Delta T_{\rm R}(F_{\rm eff})$ dependencies for individual model runs are also found to be more or less linear.  

Therefore, it is probable that the values 
of $\Delta T_{\rm R}$ for
$F_{\rm eff}\la 1.3$~(km~s$^{-1})^{-1}$ for S140 and $\la 1$~(km~s$^{-1})^{-1}$ for S199 are enhanced by some structures (processes) other than randomly distributed thermal fragments 
(e.g. gravitationally bounded compact cores, ``tangled structures'' (Hacar et al. \cite{hacar16}) or ``cloudlets'' (Tachihara et al. \cite{tachihara12})).
In order to get an unbiased estimate of $\Delta T_{\rm R}$ associated with thermal fragments we calculated linear regressions for the observed  $\Delta T_{\rm R}(F_{\rm eff})$  
dependencies in the range where dependences are nearly linear and extrapolated them to lower $F_{\rm eff}$.
Regression lines are shown in Fig.~\ref{dtr} (left).
We took the $\Delta T_{\rm R}$ values calculated from regression lines at $F_{\rm eff}$=0.7~(km~s$^{-1})^{-1}$  
as standard deviations of line temperature fluctuations produced by  
randomly distributed thermal fragments in the beam.  
Uncertainty of these estimates are assumed to be the same as uncertainty in the model calculations ($\sim 25$\%).

\section{Total Number of Fragments in the Beam}  
\label{sec:ntotal}  

Knowing $\Delta T_{\rm R}$, $T_{\rm R}$, line width, optical depth  
and kinetic temperature it is possible to get a total number  
of thermal fragments ($N_{\rm tot}$) within the telescope beam  
from equation~(\ref{eq:Ntotal}).
Kinetic temperatures ($T_{\rm KIN}$) for S140 and S199 are taken close to the estimates from  
Malafeev et al. (\cite{mal05}) and Zinchenko et al. (\cite{zin97}), respectively.  
Optical depths ($\tau$) are calculated from comparison of the HCN(1--0) and HCO$^+$(1--0) line widths with the optically thin H$^{13}$CN(1--0) and H$^{13}$CO$^+$(1--0) line widths.
For HCN(1--0) it corresponds to the $F$=2--1 hyperfine component.  
For S140 HCN(1--0) observed at IRAM-30m the $\tau$ value is taken to be the same as for the OSO-20m observations.  
Yet, this value is probably underestimated.  
Detailed model calculations (Section~\ref{sec:phys}) reproduce the IRAM-30m HCN(1--0) profile with  $\tau\sim 1$ which lead to $\sim 2.5$ times lower value of $N_{\rm tot}$.
The results are given in Table~\ref{Ntotal}.
The uncertainties of $N_{\rm tot}$ defined mainly by the $\Delta T_{\rm R}$ and $\tau$ uncertainties are at least $\sim 50$\%.  

\begin{table}[hbtp]  
\caption{Total number of thermal fragments in the beam}  
\begin{tabular}{ccccccr}\hline\noalign{\smallskip}  
Source          & Line & $\Delta T_{\rm R}(K)$ & $T_{\rm R}$(K)  & $\tau$ & $T_{\rm KIN}$(K) & $N_{\rm tot}$ \\  
\noalign{\smallskip}\hline\noalign{\smallskip}  
S 140        & HCN(1--0)    & 0.023 & 15.9(0.1)   & 0.15(0.04) & 30 & $\sim 4\cdot 10^6$ \\  
OSO-20m \\  
\noalign{\smallskip}\hline\noalign{\smallskip}  
S 140       & HCN(1--0)    &$0.017 $ & 17.6(0.1) &              &    & $\sim 10^7$ \\  
IRAM-30m \\  
\noalign{\smallskip}\hline\noalign{\smallskip}  
S 140        & HCO$^+$(1--0) & 0.013 & 21.9(0.1) & 0.42(0.04)  &    & $\sim 2\cdot 10^7$ \\  
OSO-20m \\  
\noalign{\smallskip}\hline\noalign{\smallskip}  
S199(0,0)       & HCN(1--0)        & 0.016 & 8.0(0.1)  &  0.7(0.2)    & 30 & $\sim 10^6$ \\  
IRAM-30m \\  
\noalign{\smallskip}\hline\noalign{\smallskip}  

\end{tabular}  
\label{Ntotal}  
\end{table}

\section{Physical parameters of fragments in S140}  
\label{sec:phys}  

\begin{figure}[hbtp]  
 \includegraphics[width=4.5cm,angle=-90]{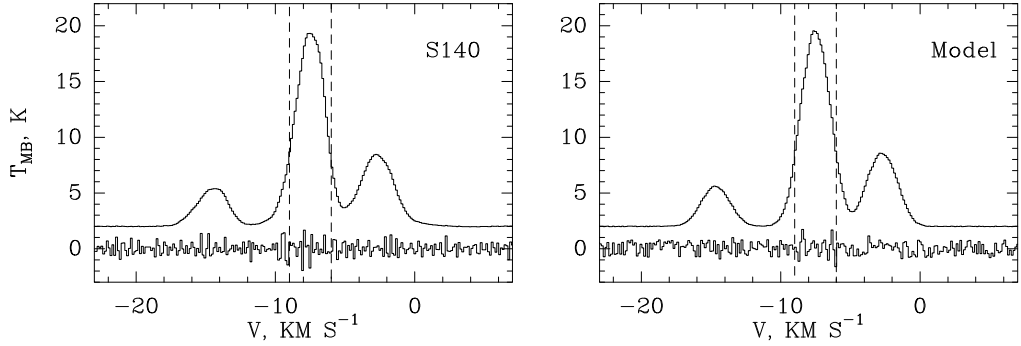}  
\caption{\small{The S140 HCN(1--0) profile observed with IRAM-30m (left)  
and the profile calculated in a framework of the 1D clumpy model (right).  
The residuals obtained by Fourier filtering with $F_{\rm eff}$=1.3~(km~s$^{-1})^{-1}$  
and multiplied by a factor of 40 are shown under each profile.  
Dashed vertical lines mark the range for which $\Delta T_{\rm R}$ is calculated.  
}}  
 \label{S140_profiles}  
 \end{figure}

We used the 1D clumpy model (see Section~\ref{sec:model}) 
for detailed modeling of the HCN(1--0) profile observed in S140 at IRAM-30m (``quiet'' component).
The model calculations reproduce very well  
the observed HCN(1--0) profile in S140 (Fig.~\ref{S140_profiles}).
Varying density and kinetic temperature of fragments, the product of the HCN abundance and a size of the cloud and velocity dispersion of relative motions of fragments it is possible to fit intensities of hyperfine components and widths.  
Varying size of fragments and its volume filling factor it is possible to fit  
standard deviation of residual temperature fluctuations ($\Delta T_{\rm R}$).
The observed and model profiles and the residuals after Fourier filtering with $F_{\rm eff}=1.3$~(km~s$^{-1})^{-1}$ multiplied by a factor of 40  
are shown in Fig.~\ref{S140_profiles}.
We added synthetic noise to the model profile with dispersion equals
to the observed one.  

The model parameters of fragments are the following:  
$T_{\rm KIN}$=30~K, number density is 1.5~10$^6$~cm$^{-3}$,
the size and the volume filling factor of fragments are $\sim 40$~a.e.  
and $\sim 0.014$, respectively.  
Optical depth of the central component ($F$=2--1) is about 1.  
The total number of fragments is $\sim 2~10^6$. This is comparable with  
the analytical estimate for HCN(1--0) in S140 IRAM-30m  (Table~\ref{Ntotal})  
if ones takes $\tau$=1.

\section{Discussion}  

New high sensitivity observations of the S140 and S199 cores confirmed an existence of residual intensity fluctuations on line profiles found previously (Pirogov \& Zinchenko \cite{pz08}, Pirogov et al. \cite{pir12}).  
Using a new method of Fourier filtering and comparing the data with the results of detailed calculations of the HCN and HCO$^+$ excitation in a framework of the clumpy model it is shown that intensity fluctuations can be associated with a large number of randomly distributed identical thermal fragments moving randomly with respect to each other with suprathermal velocities.  

The total number of fragments in the beam for S140 derived from the IRAM-30m HCN(1--0) data  
and from the OSO-20m data agree with each other if one takes $\tau$=1 for the IRAM-30m data. 
The same value derived from the HCO$^+$(1--0) data is several times higher (Table~\ref{Ntotal}).
This could imply an existence of interfragment gas of lower density which effectively absorbs the HCO$^+$ emission and reduces the corresponding $\Delta T_{\rm R}$ value.  
In order to prove this assumption model calculations in  
a framework of the model with interfragment gas are needed.
So far, the value $\sim 4~10^6$ is assumed as a reasonable estimate for total number of thermal fragments in S140.  
The uncertainty of this estimate is at least 50\%.
Following the analysis from Pirogov \& Zinchenko (\cite{pz08}) it can be shown that such fragments are unstable and short-lived density enhancements which most probably arise due to turbulence in high-mass star-forming cores.  

The difference between the new and the previous (Pirogov \& Zinchenko \cite{pz08}) results for S140 and S199 are connected mainly with a new method of estimating $\Delta T_{\rm R}$ based on the regression analysis while the previous estimates were based on Fourier filtering at arbitrary chosen value of $F_{\rm eff}=0.7$. 
The difference in line ranges for which $\Delta T_{\rm R}$ has been calculated and the difference in $\tau$ also increase the value of total number of fragments in S199.

In general, the considered model is oversimplified and the estimates obtained  
should be treated as mean values for the regions with linear sizes $\sim 0.1-0.3$~pc in the considered cores.
More realistic models should be implemented  
which would combine 3D molecular line radiation transfer in clumpy medium (e.g. Juvela \cite{j97}, Park \& Hong \cite{ph98}) with inhomogeneous turbulent cloud structures followed from modern MHD models (e.g. Haugbolle et al. \cite{h18}).
On the other hand,
a possibility to resolve by an interferometer the considered small-scale structure is not straightforward (interferometric observations usually reveal compact objects in the field of view and miss more diffuse and extended emission (see, e.g., Maud et al. \cite{maud13}, Palau et al. \cite{palau18})).
Long-time observations with high angular resolutions with single-dish telescopes seems still to be important to search for intensity fluctuations on line profiles.
An increasing sensitivity of modern receivers and implementation of new broadband spectrometers make it possible to detect residual intensity fluctuations on line profiles of various molecules in a reasonable time for different positions in objects which together with modeling results should help to get more information about their small-scale spatial and kinematic structure.

\section{Conclusions}  

Long-time observations of the S140 and S199 high-mass star-forming cores  
in the HCN(1--0) and HCO$^+$(1--0) lines were carried out.  
In order to detect intensity fluctuations on line profiles that could be due to inner small-scale structure the profiles were processed by the Fourier filtering method.  
The residual fluctuations of line radiation temperature imply an existence
a large number of randomly moving thermal fragments in the objects.  
Using analytical method a total number of fragments was calculated being $\sim 4\cdot 10^6$ for the region with linear size $\sim 0.1$~pc in S140 and $\sim 10^6$ for the region with linear size $\sim 0.3$~pc in S199.  
Physical parameters of thermal fragments in S140 were obtained from detailed modeling of the HCN excitation in a framework of the clumpy model 
including their density ($\sim 1.5~10^6$~cm$^{-3}$), size ($\sim 40$~a.e.) and volume filling factor ($\sim 0.014$).
Such fragments should be unstable and short-lived objects and are probably connected with enhanced level of turbulence in the core.

\begin{acknowledgements}  
I am grateful to the anonymous referee for critical reading of the manuscipt, valuable comments and questions which improved the paper. 
I would like to thank Olga Ryabukhina for the help in the OSO-20m observations. 
I would also like to thank Igor Zinchenko for helpful discussions.
The OSO-20m observations and the paper preparation were done under support of the RFBR grants  
(projects 15-02-06098, 16-02-00761, 18-02-00660), data procession and analysis were done  
under support of the Russian Science Foundation grant (project 17-12-01256).
\end{acknowledgements}

\label{lastpage}  


\begin{thebibliography}{99}  

\bibitem[1996]{bergin96}  
Bergin~E.~A., Snell~R.~L., Goldsmith~P.~F., 1996, ApJ, 460, 343  

\bibitem[1981]{gb81}  
Guilloteau~S., Baudry~A., 1981, A\&A, 97, 213  

\bibitem[2016]{hacar16}  
Hacar~A., Alves~J., Burkert~A., Goldsmith~P., 2016, A\&A, 591, A104  

\bibitem[2018]{h18} 
Haugbolle~T., Padoan~P., Nordlund~A., 2018, ApJ, 854, 35   

\bibitem[2008]{hirota08}  
Hirota~T., Ando~K., Bushimata~T., et al., 2008, PASJ, 60, 961  

\bibitem[1997]{j97}  
Juvela~M., 1997, A\&A, 322, 943 

\bibitem[2003]{kamegai03}  
Kamegai~K., Ikeda~M., Maezawa~H., et al., 2003, ApJ, 589, 378  

\bibitem[1986]{ks86}  
Kwan~J., Sanders~D.~B., 1986, ApJ, 309, 783  

\bibitem[2014]{lim14}  
Lim~B., Sung~H., Kim~J.~S., Bessel~M.~S., Karimov~R., 2014, MNRAS, 438, 1451  

\bibitem[2005]{mal05}  
Malafeev~S.~Yu., Zinchenko~I.~I., Pirogov~L.~E., Johansson~L.~E.~B., 2005, AstrL, 31, 239  

\bibitem[1984]{martin}  
Martin~H.~M., Sanders~D.~B., Hills~R.~E., 1984, MNRAS, 208, 35  

\bibitem[2013]{maud13}  
Maud~L.~T., Hoare~M.~G., Gibb~A.~G., et al., 2013, MNRAS, 428, 609

\bibitem[2018]{palau18} 
Palau~A., Zapata~L.~A., Rom\'an-Z\'uniga~C.~G., et al., 2018, ApJ, 855, 24 

\bibitem[1998]{ph98} 
Park~Y.-S., Hong~S.~S., 1998, ApJ, 494, 605 

\bibitem[1999]{pir99} Pirogov~L., 1999, A\&A, 348, 600  

\bibitem[2008]{pz08}  
Pirogov~L.~E., Zinchenko~I.~I., 2008, ARep, 52, 963  

\bibitem[2012]{pir12}  
Pirogov~L.~E., Zinchenko~I.~I., Johansson~L.~E.~B., Yang~J., 2012, A\&AT, 27, 475  


\bibitem[2012]{tachihara12}  
Tachihara~K., Saigo~K., Higuchi~A.~E., et al., 2012, ApJ, 754, 95  

\bibitem[2014]{tan14}
Tan~J.~C., Beltr\'an~M.~T., Caselli~P., et al., 2014, Protostars and Planets VI, Beuther~H., Klessen~R.~S., Dullemond~C.~P., Henning~Th. (eds.), University of Arizona Press, Tucson, 149

\bibitem[1996]{tauber96} 
Tauber~J.~A., 1996, A\&A, 315, 591  

\bibitem[1991]{wp91}  
White~G.~J., Padman~R., 1991,  Nature, 354, 511  

\bibitem[1997]{zin97}  
Zinchenko~I., Henning~Th., Schreyer~K.,  1997, A\&AS, 124, 385   

\end{thebibliography}
\end{document}